\documentclass[preprint]{article}

\usepackage{placeins}
\usepackage{microtype}
\usepackage{graphicx}
\usepackage{tikz}
\usetikzlibrary{bayesnet}

\usepackage{subcaption}
\usepackage{booktabs} 
\usepackage{amsmath}
\usepackage{amssymb}

\usepackage{hyperref}

\usepackage[accepted]{icml2018}

\icmltitlerunning{Stochastic Seismic Waveform Inversion using a GAN Prior}
\makeatletter
\def\runningfoot{\def\@runningfoot{}}
\def\firstfoot{\def\@firstfoot{}}
\makeatother 
\begin{document}

\twocolumn[
\icmltitle{Stochastic Seismic Waveform Inversion \\ using Generative Adversarial Networks as a Geological Prior}




\begin{icmlauthorlist}
\icmlauthor{Lukas Mosser}{ic}
\icmlauthor{Olivier Dubrule}{ic}
\icmlauthor{Martin J. Blunt}{ic}
\end{icmlauthorlist}

\icmlaffiliation{ic}{Department of Earth Science and Engineering, Imperial College London, London, United Kingdom}

\icmlcorrespondingauthor{Lukas Mosser}{lukas.mosser15@imperial.ac.uk}

\icmlkeywords{Machine Learning, ICML}

\vskip 0.3in
]



\printAffiliationsAndNotice{} 

\begin{abstract}

We present an application of deep generative models in the context of partial-differential equation (PDE) constrained inverse problems. We combine a generative adversarial network (GAN) representing an {\it a priori} model that creates subsurface geological structures and their petrophysical properties, with the numerical solution of the PDE governing the propagation of acoustic waves within the earth's interior. We perform Bayesian inversion using an approximate Metropolis-adjusted Langevin algorithm (MALA) to sample from the posterior given seismic observations. Gradients with respect to the model parameters governing the forward problem are obtained by solving the adjoint of the acoustic wave-equation. Gradients of the mismatch with respect to the latent variables are obtained by leveraging the differentiable nature of the deep neural network used to represent the generative model. We show that approximate MALA sampling allows efficient Bayesian inversion of model parameters obtained from a prior represented by a deep generative model, obtaining a diverse set of realizations that reflect the observed seismic response.
\end{abstract}

\section{Introduction}
\label{sec:introduction}
Solving an inverse problem means finding a set of model parameters that best fit observed data \cite{tarantola1984inversion,tarantola2005inverse}.
The observed data or measurements are often noisy and/or sparse and therefore lead to an ill-posed inverse problem where numerous model parameters may match the observed data \cite{kabanikhin2008definitions}. Additionally the model used to describe how the observed data are generated, the so-called forward model, may be uncertain. 
\begin{figure}[ht]
\vskip 0.2in
\begin{center}
\centerline{\includegraphics[width=\columnwidth]{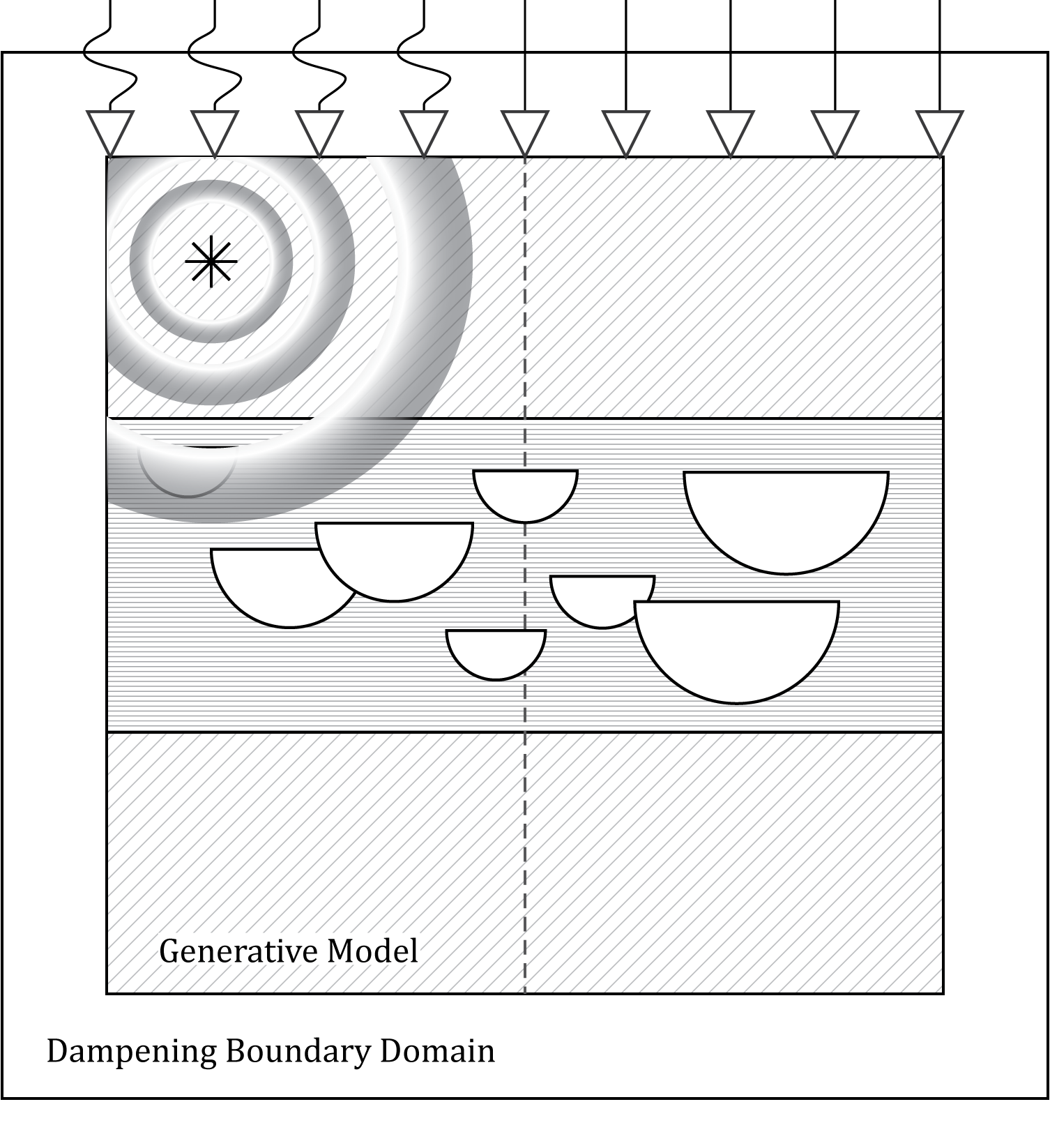}}
\caption{Computational domain for the acoustic inversion problem. Acoustic recording devices are placed on the surface ($\nabla$) and record incoming acoustic waves reflected from geological structures and emanating from an artificial source ($*$). The computational domain is embedded within a dampened boundary domain to emulate lateral and vertical dissipation of the wave-source. The generative model $G_{\theta}(\mathbf{z})$ creates the underlying spatially distributed p-wave velocity. Additional lower-dimensional constraints (dashed vertical line representing a well) can be placed on the generative model, by incorporating loss terms. The vertical axis of the computational domain has been rescaled by a factor 10 for visualization purposes.}
\label{fig:computational_domain}
\end{center}
\vskip -0.2in
\end{figure}

Based on natural observations or an understanding of the underlying data generating process we may have a pre-conception about possible or impossible states of the model parameters. We may formulate this knowledge as a prior probability distribution function (pdf) of our model parameters and use Bayesian inference to obtain a posterior pdf of the model parameters given the observations \cite{tarantola2005inverse}.

\begin{figure}[ht]
\vskip 0.2in
\begin{center}
\centerline{\includegraphics[width=\columnwidth]{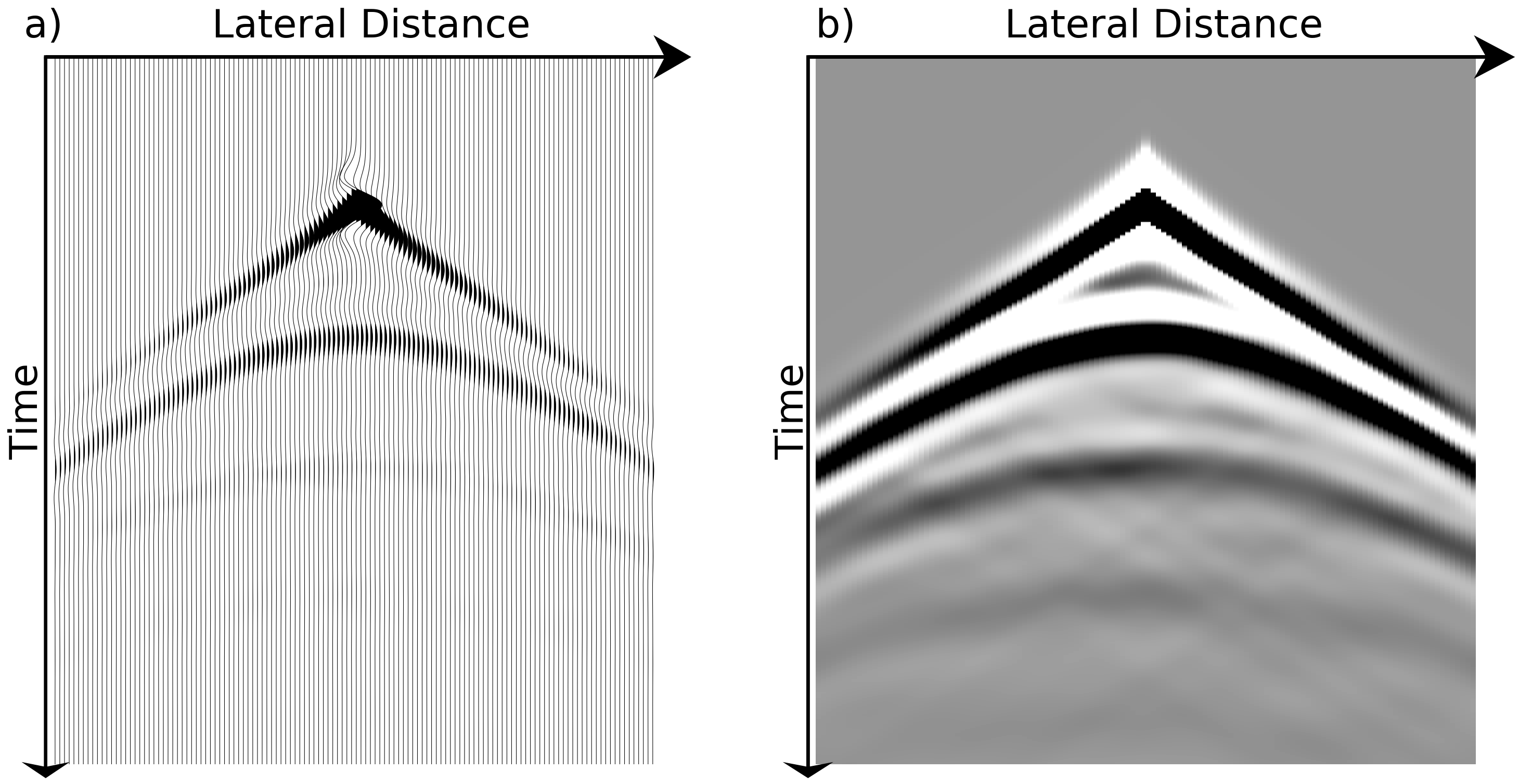}}
\caption{Recordings of acoustic waves acquired at discrete locations are spatially distributed audio signals. A seismic impulse was placed near the surface in the middle of the domain. Figure~\ref{fig:two_views}a shows the acoustic recordings as individual continuous wave forms of the measured acoustic signals. Figure~\ref{fig:two_views}b shows the same dataset of acoustic measurements  visualized as a collection of discrete pixels of an image. In this case, each sample in time is represented by a single pixel and each recorded waveform or so-called trace is represented by a column of pixels within the image.}
\label{fig:two_views}
\end{center}
\vskip -0.2in
\end{figure}

Investigating the interior structure of the earth is usually an ill-posed inverse problem \cite{tarantola2005inverse}. One method to explore the subsurface features of the earth is acoustic reflection seismic (Fig.~\ref{fig:computational_domain}) \cite{claerbout1971toward}. A number of recording devices (geophones) that record displacements are placed on the surface and a localized impulse provides an active source from which acoustic waves radiate. These waves propagate within the earth and are reflected from geological features back to the surface where geophones record the incoming signals. These recordings, seismographs\footnote{From ancient greek "Seismos": shaking, earthquake.}, represent a set of spatially distributed acoustic recordings (Fig.~\ref{fig:two_views}). The process of determining geological structures and properties of the rocks that match these data is called seismic inversion.

Seismic inversion involves modeling the physical process of waves radiating through the earth's interior. By comparing the simulated synthetic measurements to actual acoustic recordings of reflected waves we can modify model parameters and minimize the misfit between synthetic data and true measurements. The adjoint of the partial-differential equation defines a gradient of the data mismatch leading to gradient-based optimization of model parameters \cite{plessix2006review}. The set of parameters represented by the spatial distribution of the acoustic velocity of the rocks within the earth can easily exceed $10^6$ values depending on the resolution of the simulation grid and the observed data. Large three-dimensional seismic observations may require billions of parameters to be inverted for, demanding enormous computational resources \cite{akcelik2003high}.

Direct observations of the earth's interior are extremely difficult to obtain. Bore-holes may have been drilled for hydrocarbon exploration/development or hydrological measurements. These represent a quasi one-dimensional source of information of an extremely localized nature. Typical bore-hole sizes are on the order of 10s of centimeters in diameter whereas the resolution of seismic observations is usually on the order of 10s of meters. 

Nevertheless, we can deduce prior knowledge of the earth's interior from surface observations and exposed geological features. The principle of gradualism \cite{hutton1788x} states that processes governing the earth's surface today are the same processes that controlled the deposition and erosion of ancient geological features now buried deep within the earth.
This geological knowledge can be incorporated into prior distributions of physical properties of rocks, such as the acoustic p-wave velocity, or into geological features such as geological facies and fault distributions within the earth.

Efficient parameterizations \cite{akcelik2002parallel,boehm2016wavefield} that allow a dimensionality reduced representation of the high-dimensional parameter space of possible models may allow for improved inversion techniques. Due to the high computational cost incurred by inversion \cite{modrak2015,akcelik2003high}, probabilistic ensembles of models that match observed data are rarely generated and often only a single model that satisfies pre-defined quality criteria is created and used for interpretation and decision making processes.

We parameterize a set of geological features by a deep generative model that creates stochastic realizations of possible model parameters. The probabilistic distribution of model parameters is parameterized by a lower-dimensional set of normally distributed latent variables. Combined with a generative deep neural network this represents a differentiable prior on the possible model parameters. We combine this differentiable generative model with the numerical solution of the acoustic wave equation to produce synthetic acoustic observations of the earth's interior \cite{louboutin2017full}. Using the adjoint method \cite{plessix2006review}, we compute a gradient with respect to model parameters not directly in the high-dimensional model space, but in the much smaller set of latent variables. These gradients are required to perform a Metropolis-adjusted Langevin (MALA) sampling of the posterior of the model parameters given the observed seismic data. Performing MALA sampling allows us to obtain a diverse ensemble of model parameters that match the observed seismic data. Additional constraints on the generative model, such as information located at an existing bore-hole, are readily incorporated and included in the MALA sampling procedure. 

We summarize our contributions as follows:
\begin{itemize}
\item We combine a differentiable generative model controlled by a set of latent variables with the solution of a PDE-constrained numerical solution of a physical forward problem.
\item We use gradients obtained from the adjoint method to perform MALA sampling of the posterior in the lower-dimensional set of latent variables.
\item We illustrate the proposed inversion framework using a simple seismic inversion problem and evaluate the resulting ensemble of model parameters.
\item The framework allows integration of additional information, such as the knowledge of geological facies along one-dimensional vertical bore-holes.
\item The proposed approach may readily be extended to a number of inverse problems where gradients of the objective function with respect to input parameters can be calculated.
\end{itemize}

\section{Related Work}
\citet{tarantola1987inverse} cast the geophysical seismic inversion problem in a Bayesian framework.
\citet{mosegaard1995monte} presented a general methodology to perform probabilistic inversion using Monte Carlo sampling. They sampled the prior distribution of model parameters using a random walk and subsequently sampled from their posterior using a Metropolis rule. In a similar manner, \citet{sen1996bayesian} evaluated the use of Gibbs sampling to obtain {\it a posteriori} model parameters and evaluate parameter uncertainties. \citet{mosegaard1998resolution} showed that the general Bayesian inversion approach of \citet{mosegaard1995monte} could not only provide model parameter covariances but also gave information on the ability to resolve geological features.
Geostatistical models allowed spatial relationships and dependencies of the petrophysical parameters to be modeled and incorporated into a stochastic inversion framework~\cite{bortoli1993constraining,haas1994geostatistical}. \citet{buland2003bayesian} have developed an approach to perform Bayesian inversion for elastic petrophysical properties in a linearized setting.

In the case of geophysical inverse-problems computation of the solution to the forward problem is highly expensive. 
Therefore, computationally efficient approximations to the full solution of the wave-equation may allow efficient solutions to complex geophysical inversion problems. Neural networks have been shown to be universal function approximators \cite{hornik1989multilayer} and as such lend themselves as possible proxy-models for solutions to the geophysical forward and inverse problem. 

The early work by \citet{roth1994neural} presents an application of neural networks to invert from an acoustic time-domain of seismic amplitude responses to a depth profile of acoustic velocity in a supervised setting. They used pairs of synthetic data and velocity models to train a multi-layer feed-forward neural network with the goal of predicting acoustic velocities from recorded data only. They showed that neural networks can produce high resolution approximations to the solution of the inverse problem based on representations of the input model parameters and resulting synthetic waveforms alone. In addition, they showed that neural networks can invert for geophysical parameters in the presence of significant levels of acoustic noise.

Representing the geophysical model parameters at each point in space quickly leads to a large number of model parameters especially in the case of three-dimensional problems. \citet{2017arXiv171209685B} represented the spatially varying coefficients that govern the solution of a partial differential equation (PDE) by a neural network. The neural network acts as an approximation to the spatially varying coefficients characterized by the weights of the neural network. The weights of the individual neurons are modified by leveraging the adjoint-state equation in the reduced-dimensional space of network-parameters rather than at each spatial location of the computational grid. 

\citet{hansen2017efficient} replaced the solution of the partial differential equation by a neural network allowing fast computation of forward models and facilitating a solution to the inversion problem by Monte-Carlo sampling. \citet{araya2018deep} used deep neural networks to perform a mapping between seismic features and the underlying p-wave velocity domain; they validated their approach based on synthetic examples. 
Recently, a number of applications of deep generative priors have been presented in the context of computer vision for linear \cite{linearinverse} and bilinear \cite{bilinearinverse} inverse problems, as well as compressed sensing \cite{compressedgan}. For more general geophysical inverse problems, \citet{2017arXiv170804975L} have used a  variational auto-encoder to create geological models used for hydrological inversion. Inversion was performed using an adapted Markov-Chain Monte-Carlo (MCMC) \cite{laloy2012high} algorithm where the generative model was used as an unconditional prior to sample hydrological model parameters.

\citet{model2seismic} used a GAN with cycle-constraints (cycle-GAN) \cite{cyclegan} to perform seismic inversion formulating the inversion task as a domain-transfer problem. Their work used a cycle-GAN to map between the seismic amplitude domain and p-wave velocity models. Due to the p-wave velocity models and seismic amplitudes being represented as a function of depth, rather than depth and time respectively, this approach lends itself to perform stratigraphic inversion, where a pre-existing velocity model is used to perform time-depth conversion of the seismic amplitudes. \citet{modelorder} used a quasi-Newton method to optimize model parameters in the latent-space of a pre-trained GAN for a synthetic salt-body benchmark dataset.

\section{Problem Definition}
\label{sec:problem}
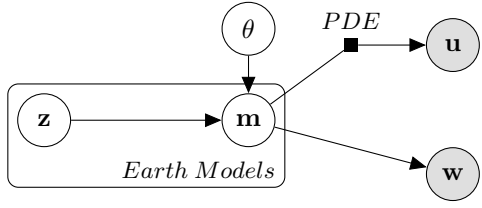
\begin{figure}
  \begin{center}
    \begin{tabular}{cc}
    \begin{tikzpicture}

      \node[obs]                               (U) {$\mathbf{u}$};
      \node[obs, below=of U]                   (W) {$\mathbf{w}$};
	  \node[latent, left=of U, xshift=-1.cm, yshift=-1cm]  (M) {$\mathbf{m}$};
      \node[latent, left=of M, xshift=-1.cm]  (Z) {$\mathbf{z}$};
 	  \node[latent, above=of M, xshift=0cm, yshift=-0.5cm]  (T) {$\mathbf{\theta}$};     
      \edge {Z} {M};
      \edge {T} {M};
      \edge {M} {W};
	  \factor[left=of U, xshift=-0.5cm] {M-U} {above:$PDE$} {M} {U} ; %
      \plate {pde} {(Z)(M)} {$Earth \ Models$} ;
    \end{tikzpicture}
    \end{tabular}
  \end{center}
  \caption{Graphical model of the geological inversion problem. The set of possible earth models are represented by a generative model with parameters $\theta$ (the parameters of the generator $\mathbf{m} \sim G_{\theta}(\mathbf{z})$). We obtain model observations of the acoustic waves $\mathbf{u}$ via the deterministic partial-differential equation (PDE) as well as partial observation of the latent model parameters $\mathbf{m}$ from local information at e.g. bore-holes $\mathbf{w}$.}\label{fig:graphical}
\end{figure}
\subsection{Bayesian Inversion}\label{sec:mala}
In the Bayesian framework of inverse problems we aim to find the posterior of latent variables $p(\mathbf{z}|\mathbf{d_{obs}})$ given the observed acoustic reflection data $\mathbf{d}_{obs}$ (Fig.~\ref{fig:graphical}). The joint probability of the latent variables $\mathbf{z}$ and observed seismic data $\mathbf{d}$ is
\begin{equation}\label{equ:simplified_joint}
p(\mathbf{z}, \mathbf{d}) = p(\mathbf{d}|\mathbf{z})p(\mathbf{z})
\end{equation}
Furthermore, by applying Bayes rule, we define the posterior over the latent variables $\mathbf{z}$ given the observed seismic data $\mathbf{d}$
\begin{equation}\
p(\mathbf{z}|\mathbf{d}) = \frac{p(\mathbf{d}|\mathbf{z})p(\mathbf{z})}{p(\mathbf{d})} \propto p(\mathbf{d}|\mathbf{z})p(\mathbf{z})
\end{equation}
We represent the observed data by 
\begin{equation}
\mathbf{d} = S(\mathbf{m}) + \mathbf{\epsilon}
\end{equation}
where $S(\mathbf{m})=S(m(\mathbf{x}))=S(G_{\theta}(\mathbf{z}))$, denoting the spatial model coordinates as $\mathbf{x}$, and $S$ is the seismic forward modeling operator. The geological facies $\mathbf{m}^{facies}$, the p-wave velocity $\mathbf{m}^{V_p}$, and the rock density $\mathbf{m}^{\rho}$ represent the set of model parameters $\mathbf{m}$. The model parameter $\mathbf{m}^{facies}$ represents the probability of a geological facies to occur at a spatial location $\mathbf{x}$, whereas  $\mathbf{m}^{V_p}$ and $\mathbf{m}^{\rho}$ represent spatial distributions of rock properties. We assume a normally distributed noise term $\mathbf{\epsilon}$.

The aim is to find samples of the posterior $\mathbf{z}_t\sim p(\mathbf{z}|\mathbf{d})$. We reformulate the approach using an approximate Metropolis-adjusted Langevin sampling rule (MALA-approx) \cite{roberts1996exponential,roberts1998optimal,2016arXiv161200005N}
\begin{equation}
\mathbf{z}_{t+1}=\mathbf{z}_{t}+\epsilon_1 \nabla \log~p(\mathbf{z}_{t}|\mathbf{d}) + \mathcal{N}(0, \epsilon_{2})
\end{equation}
Rewriting the log-likelihood of the data given the latent variables in terms of the $L_2$-norm 
$\log~p(\mathbf{d}|\mathbf{z}_{t}) \propto \|S(G_{\theta}(\mathbf{z}_{t}))-d^{obs}\|_2$ leads to the proposal rule of the MALA approximation \cite{2016arXiv161200005N}
\begin{equation}\label{equ:mala}
\begin{split}
\mathbf{z}_{t+1}=(1-\lambda)\mathbf{z}_{t} +\epsilon_1 \frac{\partial \|S(G_{\theta}(\mathbf{z}_{t}))-d^{obs}\|_2}{\partial G_{\theta}(\mathbf{z}_{t})} \frac{\partial G_{\theta}(\mathbf{z}_{t})}{\partial \mathbf{z}_{t}} + \\ \mathcal{N}(0, \epsilon_{2})
\end{split}
\end{equation}
Using this sampling approach requires gradients of the data mismatch with respect to model parameters of the forward problem, which are obtained by the adjoint-state method which will be presented in the following section. The gradients of the model parameters $\frac{\partial G_{\theta}(\mathbf{z}_{t})}{\partial \mathbf{z}_{t}}$ with respect to the latent variables are obtained by traditional neural network backpropagation.

We follow the MALA step-proposal algorithm using MALA parameters $\lambda=10^{-5},~\epsilon_1=10^{-1},~\epsilon_2=2\epsilon_{1}$ for all our simulations \cite{stepsize}. To obtain valid samples of the posterior we furthermore anneal the step size from the initial value of $\epsilon_1=10^{-1}$ to $\epsilon_1=10^{-5}$ over 100 iterations.

Where lower-dimensional information is available, such as at bore-holes, the geological models should honor not only the seismic response but also this additional information. In this study we additionally evaluate the possibility to obtain samples of the posterior that reflect observed geological facies indicators $\mathbf{m}^{facies}$ at a one-dimensional bore-hole.  When including bore-hole information the step-proposal corresponds to
\begin{equation}\label{equ:mala_extended}
\begin{split}
\mathbf{z}_{t+1}=(1-\lambda)\mathbf{z}_{t} +\epsilon_1 \frac{\partial \|S(G_{\theta}(\mathbf{z}_{t}))-d^{obs}\|_2}{\partial G_{\theta}(\mathbf{z}_{t})} \frac{\partial G_{\theta}(\mathbf{z}_{t})}{\partial \mathbf{z}_{t}} + \\ \epsilon_3 \frac{\partial \log~p(\mathbf{m}^{facies}=\mathbf{m}^{facies}_{well}|\mathbf{z}_t)}{\partial \mathbf{z}_{t}} + \mathcal{N}(0, \epsilon_{2})
\end{split}
\end{equation}
where we seek to obtain samples of the posterior given the observed seismic data $d^{obs}$ and geological facies at the wells $\mathbf{m}^{facies}_{well}$. 

The additional term $\log~p(\mathbf{m}^{facies}=\mathbf{m}^{facies}_{well}|\mathbf{z}_t)$ in equation~\ref{equ:mala_extended} is equal to the log-likelihood of the facies probability, as derived from the generator, calculated at the bore-hole. In all our simulations we set $\epsilon_{3}=100$ when bore-hole data is incorporated.
\subsection{Adjoint-State Method}\label{sec:adjoint}
We perform numerical solution of the time-dependent acoustic wave equation given a set of model parameters
\begin{equation}\label{equ:wave_equation}
\begin{split}
L(\mathbf{u}, \mathbf{m}^{V_p}) = \frac{1}{m^{V_p}(\mathbf{x})^{2}}\frac{d^2{u(\mathbf{x}, t)}}{d{t}^2}-\Delta{u(\mathbf{x}, t)} \\
+\eta\frac{d{u(\mathbf{x}, t)}}{d{t}}=q(\mathbf{x}, \mathbf{x_s}, t)
\end{split}
\end{equation}
where $u(\mathbf{x}, t)$ is the unknown wave-field and $m^{V_p}(\mathbf{x})$ is the acoustic p-wave velocity. The dampening term $\eta\frac{d{u(\mathbf{x}, t)}}{d{t}}$ prevents reflections from domain boundaries and ensures that waves dissipate laterally.

Time-dependent source wavelets $q(\mathbf{x}, \mathbf{x_s}, t)$ are introduced at source locations $\mathbf{x}_s$. We emulate the seismic acquisition process by placing acoustic receivers that record the incoming wave-field at the top edge of the simulation domain. 
While receiver locations remain fixed, the number of sources is varied to image a number of parts of the subsurface domain. 

We introduce a compact notation of equation~\ref{equ:wave_equation}
\begin{equation}\label{equ:compact}
F(\mathbf{u}, \mathbf{m}^{V_p})=L(\mathbf{u}, \mathbf{m}^{V_p})-q_s=0
\end{equation}
which we refer to as the forward-problem.

To perform sampling according to the MALA algorithm presented in equation~\ref{equ:mala}, we seek to obtain a gradient of the following functional
\begin{equation}\label{equ:minimize}J(m^{V_p}(\mathbf{x}))=\sum^{n_{sources}}_{i=1}{\|d^{pred}(m^{V_p}(\mathbf{x})), q_i)-d^{obs}\|^2}
\end{equation}
where $d^{pred}$ and $d^{obs}$ are the predicted and observed acoustic observations respectively.

We  augment the functional $J(m^{V_p}(\mathbf{x}))$ by forming the Lagrangian
\begin{equation}\label{equ:lagrangian}
L(\mathbf{m}^{V_p}, ~\mathbf{u}, ~\mathbf{\lambda})=J(\mathbf{m}^{V_p})-\langle\mathbf{\lambda}, ~F(\mathbf{u}, \mathbf{m}^{V_p})\rangle
\end{equation}

Differentiating $L(\mathbf{m}^{V_p}, ~\mathbf{u}, ~\mathbf{\lambda})$ with respect to $\lambda$ leads to the state equation~\ref{equ:wave_equation}, but differentiation with respect to the acoustic wave-field $\mathbf{u}$ leads to the adjoint state equations \cite{plessix2006review}:
\begin{equation}
(\frac{\partial{F(\mathbf{u}, \mathbf{m}^{V_p})}}{\partial{\mathbf{u}}})^T~\mathbf{\lambda}=(d^{pred}-d^{obs})
\end{equation}
showing that we obtain a similar back-propagation equation as that used to derive gradients in neural networks \cite{lecun1988theoretical}: the data mismatch is backpropagated thanks to a linear equation in the adjoint state vector $\mathbf{\lambda}$. By differentiating the Lagrangian in equation~\ref{equ:lagrangian} with respect to $m(\mathbf{x})$ we obtain
\begin{equation}\label{equ:adjoint_gradient}
\frac{\partial{J}}{\partial{\mathbf{m}^{V_p}}}=-\langle\mathbf{\lambda}, ~ \frac{\partial{F(\mathbf{u}, ~\mathbf{m}^{V_p})}}{\partial{\mathbf{m}^{V_p}}}\rangle=\frac{\partial \|S(G_{\theta}(\mathbf{z}_{t}))-d^{obs}\|_2}{\partial G_{\theta}(\mathbf{z}_{t})} 
\end{equation}
which is equivalent to the gradient required to perform MALA sampling of the posterior distribution of latent variables (Eq.~\ref{equ:mala}).

We perform numerical solution of the acoustic wave-equation and the respective adjoint computation using the domain-specific symbolic language Devito \cite{2016arXiv160808658K,louboutin2017full}. Numerical solution is performed using a fourth-order finite-difference scheme.

\begin{figure*}
\vskip 0.2in
\begin{center}
\centerline{\includegraphics[width=\textwidth]{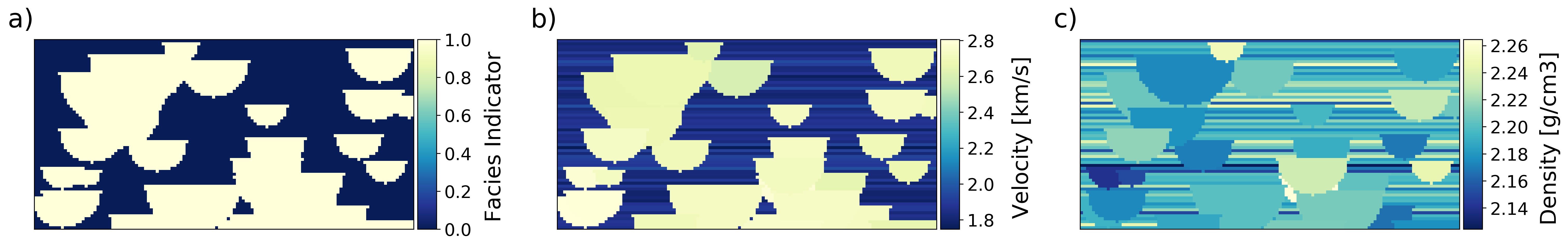}}
\caption{Overview of the object-based model realization used as ground truth for evaluating the inversion procedure. Geological facies (a) distinguish between river channel bodies (light) and shale (dark). Acoustic p-wave velocity $V_p$ (b) and rock density $\rho$ (c) are constant within river-channels and vary by layer within shale.}
\label{fig:ground_truth}
\end{center}
\vskip -0.2in
\end{figure*}

\section{Generative Model}\label{sec:generative_model}
We parameterize the model parameters of the acoustic wave equation (Eq.~\ref{equ:wave_equation}),
by a differentiable generative model $G_{\theta}(\mathbf{z})$. We use a generative model to sample realizations of spatially varying model parameters $m(\mathbf{x})\sim G_{\theta}(\mathbf{z})$. These realizations are obtained by sampling a number of latent variable vectors $\mathbf{z}$. The associated model representations represent the {\it a priori} knowledge about the spatially varying properties of the geological structures in the subsurface.

We model the prior distribution of the spatially varying model parameters $m(\mathbf{x})$ (Sec.~\ref{sec:mala}) by a generative adversarial network (GAN) \cite{goodfellow2014}. GANs represent a generative model where the underlying probability density function is implicitly defined by a set of training examples. To train GANs two functions are required: a generator $G_{\theta}(\mathbf{z})$ and a discriminator $D_{\omega}(\mathbf{m})$. The role of the generator is to create random samples of an implicitly defined probability distribution that are statistically indistinguishable from a set of training examples. The discriminator's role is to distinguish real samples from those created by the generator. Both functions are trained in a competitive two-player min-max game where the overall loss is defined by:
\begin{equation}
\label{equ:minmax}
\begin{split}
\min_{\mathbf{\theta}} \max_{\mathbf{\omega}}\{\mathbb{E}_{\mathbf{m}\sim p_{\mathbf{m}}}[\log \ D_{\mathbf{\omega}}(\mathbf{m})] \\ + \mathbb{E}_{\mathbf{z}\sim p_{\mathbf{z}}}[\log~(1-D_{\mathbf{\omega}}(G_{\mathbf{\theta}}(\mathbf{z})))]\}
\end{split}
\end{equation}
Due to the opposing nature of the objective functions, training GANs is inherently unstable and finding stable training methods remains an open research problem. Nevertheless, a number of training methods have been proposed that allow more stable training of GANs. In this work we use a so-called Wasserstein-GAN \cite{arjovsky2017wasserstein,gulrajani2017improved}, that seeks to minimize the Wasserstein distance between the generated and real probability distribution. We use a Lipschitz penalty term proposed by \citet{2017arXiv170908894P} to stabilize training of the Wasserstein-GAN. The full objective function including the gradient penalty term is given by:
\begin{equation}\label{eq:wgan}
\begin{split}
  \min_{\mathbf{\theta}} \max_{\mathbf{\omega}}\mathbb{E}_{\mathbf{z}\sim p_{\mathbf{z}}}[D_{\mathbf{\omega}}(G_{\mathbf{\theta}}(\mathbf{z}))]  -\mathbb{E}_{\mathbf{m}\sim p_{\mathbf{m}}}[D_{\mathbf{\omega}}(\mathbf{m})] \\ 
  +\lambda_{LP} \mathbb{E}_{\hat m \sim \tau}[\left( \max \left \{ 0 , \|\nabla D_{\mathbf{\omega}} (\hat m)\| -1 \right \} \right)^2]
\end{split}
\end{equation}
where $\tau$ is a statistical distribution controlling a linear combination between a real and generated sample \cite{2017arXiv170908894P}.

In our work we set $\lambda_{LP}=200$ to train the generative model. We represent both the generator and discriminator\footnote{In the Wasserstein-GAN literature the discriminator is also termed a "critic".} function by deep fully convolutional neural networks (see Appendix Table~\ref{tab:gan_architecture}). The generator uses a number of convolutional layers followed by so-called pixel-shuffle transformations to create output domains \cite{pixelshuffle}. 

The latent vector is parametrized as a multivariate standardized normal distribution:
\begin{subequations}
\begin{eqnarray}
\mathbf{z} \sim \mathcal{N}(0, \mathbf{I})^{50 \times 1 \times 2}
\label{equ:noise_prior} \\
 G_{\mathbf{\theta}}: \mathbf{z} \rightarrow \mathbb{R}^{3 \times 64 \times 128}
\label{equ:generator_mapping}
\end{eqnarray}
\end{subequations}
Due to the geological properties represented in our dataset, namely, geological facies indicators $\mathbf{m}^{facies}$, acoustic p-wave velocity $\mathbf{m}^{V_p}$ and density $\mathbf{m}^{\rho}$, the generator must output three data channels. We represent the geological facies as a probability of a spatial location belonging to a shale or sandstone facies. To facilitate numerical stability of the GAN training process we apply a hyperbolic tangent activation function and convert to a probability $\mathbf{m}^{facies}$ for subsequent computation (Eq.~\ref{equ:mala_extended}). The acoustic p-wave velocity $V_p$ is represented by a Gaussian distribution within each modeled geological facies. We apply a hyperbolic tangent activation function to model the output distribution of the p-wave model parameters $\mathbf{m}^{V_p}$. The rock density $\mathbf{m}^{\rho}$ follows a Gaussian distribution and a soft-plus activation function is used to ensure positive values of rock density (Appendix~\ref{sec:appendix}). In this study, only the facies indicator $\mathbf{m}^{facies}$ and acoustic p-wave velocity $\mathbf{m}^{V_p}$ are used in the inversion process.

The generator-discriminator pairing is trained on the set of training images described in section~\ref{sec:dataset}. After training, the generator $G_{\theta}(\mathbf{z})$ and the forward modeling operator $S(\mathbf{m})$ are arranged in a fully differentiable computational graph.  To accommodate the sources and receivers of the acoustic forward modeling process described in section~\ref{sec:problem}, we pad the output of the generator by a domain of constant p-wave velocity.

\section{Dataset}
\label{sec:dataset}
Geological structures in the earth's interior often closely resemble features observed at the surface. As sediments accumulate over time underlying rock is buried and exposed to high temperatures and pressures, deforming and compacting sediments. Ancient river systems often represent pathways for fluids with high storage capacity and permeability and are therefore common targets for hydrocarbon exploration. 

To demonstrate the proposed inversion method we will use a model of a fluvial-dominated system consisting of highly porous sandstones embedded in a fine grained shaly material. Object-based models are commonly used to model such geological systems \cite{deutsch1996hierarchical}. They represent the fluvial environment as a set of randomly located geometric objects following various size, shape and property distributions. We train a set of GANs on a dataset of ten thousand realizations of two-dimensional cross-sections of fluvial object-based models. 

\begin{figure*}[ht]
\vskip 0.2in
\begin{center}
\centerline{\includegraphics[width=\textwidth]{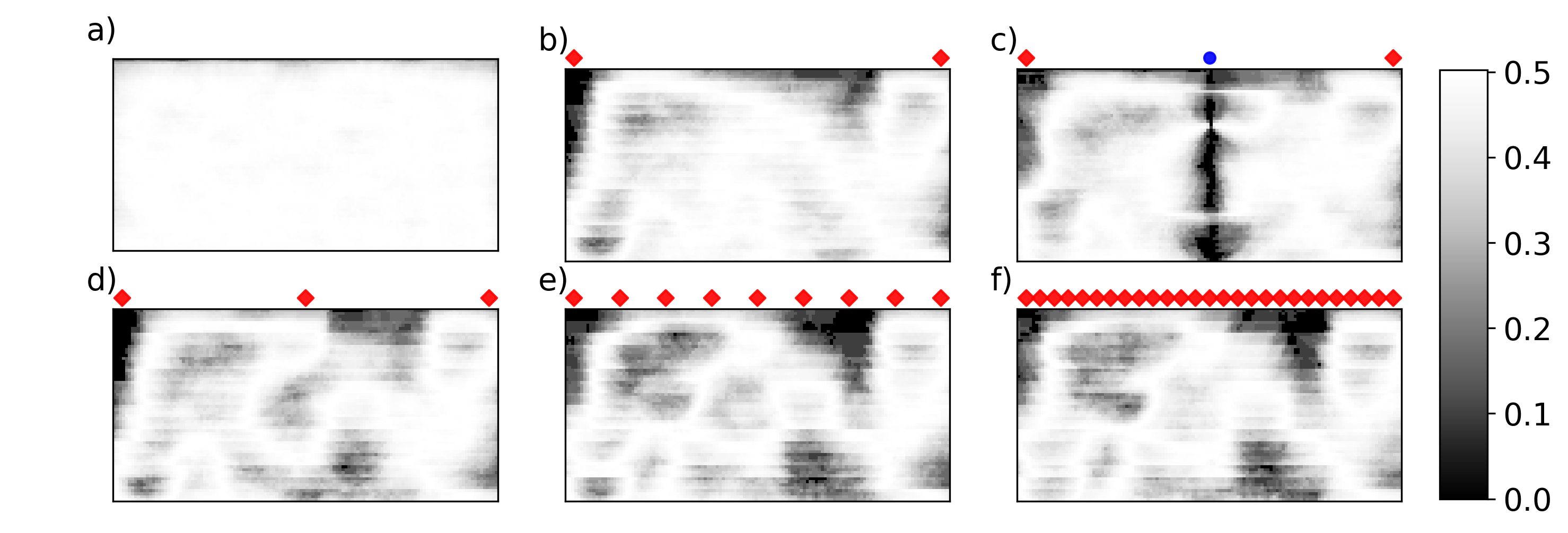}}
\caption{Pixel-wise standard deviation of an ensemble of 200 models sampled unconditionally from the prior (a) represented by the generator function $G_{\theta}(\mathbf{z})$. Posterior ensemble of geological indicator variables matched to the seismic representation of the ground truth image shown in figure \ref{fig:ground_truth} for (b) 2 sources, (c) 2 sources and a single bore-hole, (d) 3 sources, (e) 9 sources, (f) 27 sources. Source locations are indicated by red diamonds and the bore-hole location by a blue circle.}\label{fig:errorbars}
\end{center}
\vskip -0.2in
\end{figure*}

\begin{figure*}[h]
\vskip 0.2in
\begin{center}
\centerline{\includegraphics[width=\textwidth]{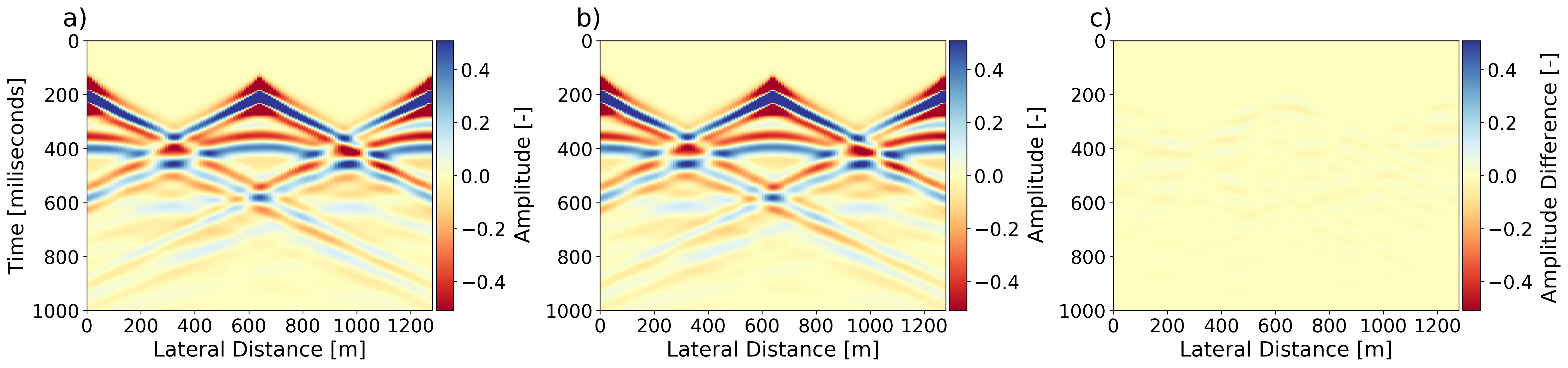}}
\caption{Comparison of the seismic waveform based on the ground truth acoustic velocity (a) compared to the waveform of an inverted model with three seismic sources (b). The difference (c) of the two waveforms has a relative error $<\epsilon_{seismic}=10\%$. Colormaps are scaled based on one standard deviation in amplitudes of the ground truth wave-form (a).}
\label{fig:mismatch}
\end{center}
\vskip -0.2in
\end{figure*}
The individual cross-sections are created with an object-based model, where half-circle sand-bodies follow a uniform width distribution and their p-wave velocity and density are constant for each channel-body. The locations of individual channel-bodies are determined by a uniform distribution in spatial location. The fine-grained material surrounding the river-systems is made of layers of single-pixel thickness where each layer has a constant value of acoustic p-wave velocity and density which varies randomly but marginally from one layer to another. We use a binary indicator variable to distinguish the two facies regions, river-channel vs shale-matrix. The ratio of how much of a given cross-section is filled with river-channels compared to the overall area of the geological domain is a key property in understanding the geological nature of these structures. This ratio follows a uniform distribution in our dataset and river-channels are placed at random until a cross-section meets the necessary ratio criterion.

A total of ten thousand training images were created as a training set for the GAN. A further five thousand images were retained as a test set to evaluate the inversion technique. While training the generative model outlined in section \ref{sec:generative_model} we monitor image quality and output distributions for each of the modeled properties.

Figure \ref{fig:ground_truth} shows a comparison of the distribution of the three-modeled properties, geological facies indicator, acoustic p-wave velocity and rock density for the ground-truth example that was used to emulate a set of obtained measured seismic data.

\section{Results}
We evaluate the proposed method of inversion by sampling a set of latent variables $\mathbf{z}$ determining the output of the generative model $G_{\theta}(\mathbf{z})$ (Sec.~\ref{sec:dataset}, Fig.~\ref{fig:ground_truth}). First, we evaluate the generative model as a prior for representing possible earth models and generating $N=100$ unconditional samples.
The resulting ensemble shows high variability in geological structures and closely matches the distribution of geophysical properties defined in the training dataset (Fig.~\ref{fig:ground_truth}a). 

Two cases of inversion are considered: inversion for the acoustic p-wave velocity $V_p$ and combined inversion of acoustic velocity and of geological facies along a bore-hole. For all tests we perform inversion using the MALA-approx scheme and accept inverted models with a relative seismic error $\epsilon_{seismic}$ of less than 10\%. For the additional bore-hole constraint we require an accuracy above 95\% of geological facies to be accepted as a valid inverted sample. While lower relative errors in seismic mismatch and bore-hole accuracy can be achieved, evaluating the forward problem and adjoint of the partial differential equation comes at high computational cost and therefore a cost-effectiveness trade-off was necessary. 
\begin{figure}[ht]
\vskip 0.2in
\begin{center}
\centerline{\includegraphics[width=\columnwidth]{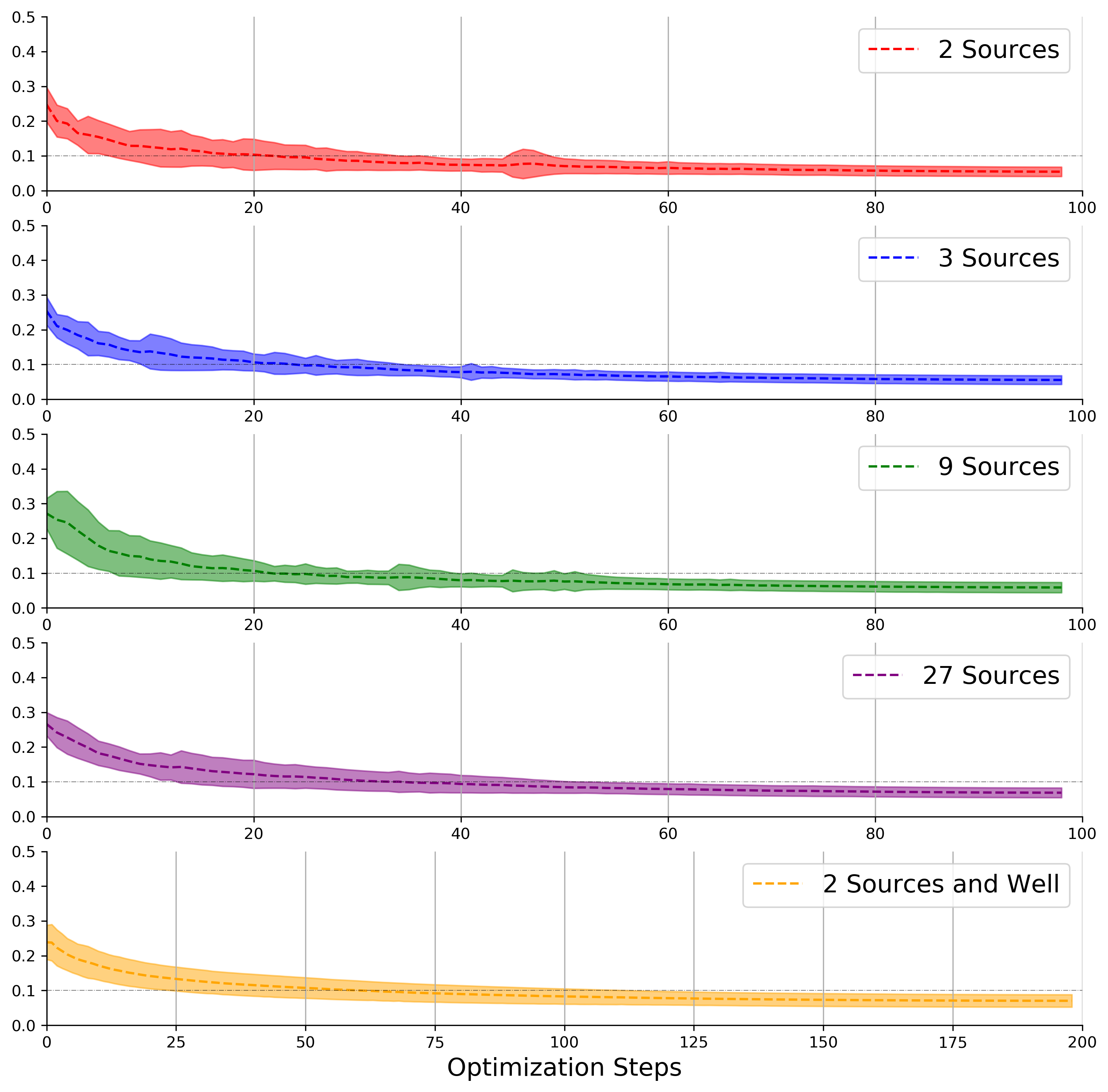}}
\caption{Comparison of the relative error in seismic data mismatch. Shaded regions indicated $\pm \ \sigma$ in the relative error. We perform 100 iterations of MALA-approx to obtain samples of the posterior given seismic observations only, and 200 iterations where bore-hole information is included. The step-size of the Markov-Chain was annealed to very small values leading to a stabilization of the loss at the end of the sampling procedure.}
\label{fig:convergence}
\end{center}
\vskip -0.2in
\end{figure} 

For the first case of seismic inversion without model constraints we perform simulations where the number of acoustic sources are increased. Fewer acoustic sources means that less of the domain is properly imaged, leading to high uncertainty in areas where no incoming waves have been reflected and recorded by the receivers on the surface. Figure~\ref{fig:errorbars} shows a comparison of the evaluated inverted models for $N=100$ inverted earth models. Acoustic sources are shown by red diamonds. The acoustic sources and 128 receivers are equally spaced across the top edge of the domain.

In Figure~\ref{fig:errorbars} (b, d-f) we show the pixel-wise standard deviation of 100 inverted models for an increasing number of acoustic sources (2 sources up to 27 sources). As the total number of acoustic sources increases we obtain a lower standard deviation for the resulting model ensembles. In the case of two acoustic sources (Fig.~\ref{fig:errorbars}b) we find that close to the sources there is small variation amongst the inverted models (dark shades) whereas the central area where no acoustic source has been placed shows a very high degree of variation. This is confirmed by the three source case where a central acoustic source has been placed in addition to the sources on the borders of the domain. Lower variability in the inverted ensemble and structural features can be observed. This correlates well with the Bayesian interpretation of the inverse problem; where acoustic sources allow the subsurface to be imaged we arrive at a low standard deviation in the posterior ensemble of geological models, whereas within regions that are only sparsely sampled by the acoustic sources we expect the prior, the unconditional generative model, to be more prevalent, leading to a higher variability of geological features. As expected, when we increase the number of sources, we find overall smaller variability in the resulting ensemble of inverted earth models. We observe only marginal reduction in variability between the case of nine and twenty-seven sources (Fig.~\ref{fig:errorbars}e-f) as we reach the limits of resolution of the forward problem.

In the case where lower-dimensional information such as a bore-hole was included as an additional objective function constraining the generative model  (Fig.~\ref{fig:errorbars}b), we find a lower standard deviation around this bore-hole. The standard deviation along the well is close to zero due to the per-realization 95\% accuracy constraint. Furthermore there is a region of influence where the bore-hole constrains lateral features such as channel bodies. This is shown by channel shaped features of low standard deviation at the top and bottom of the domain. Comparison with the ground truth image (Fig.~\ref{fig:ground_truth}a) shows that two channel bodies can be found along the one-dimensional feature. 

For each generated realization we have recorded the relative error in the seismic data mismatch (Fig.~\ref{fig:mismatch}) as a function of the required iterations to perform the inversion as well as the latent variable vectors $\mathbf{z}$ at each MALA sampling iteration. In practice we find that performing 100 iterations of the MALA approximation leads to a sufficient reduction in the seismic likelihood and as the step-size is reduced linearly, the seismic data mismatch stabilizes at relative errors of 5-7\%. When bore-hole constraints are included (Fig.~\ref{fig:convergence} bottom) we need to perform 200 MALA-approx iterations to find inversions that respect seismic observations and bore-hole features.  

Due to the fact that modern full-waveform inversion (FWI) methods come at very high computational cost for two and possibly three-dimensional inversions, a small number of required iterations is a necessity. In further tests, reducing the number of iterations of the MALA approximation or simply minimizing the seismic loss by gradient descent, as performed by \citet{modelorder}, allows convergence to small relative errors but this approach has been shown to lead to reduced sample diversity \cite{2016arXiv161200005N}.

Due to the differentiable nature of the generative model, whose parameters are kept constant for inversion, it may be possible to deduce much more efficient implementations that do not require computation of the gradient of the high-dimensional model parameters $\mathbf{m}$ but rather evaluate gradients direct with respect to the latent variables.  

Using a probabilistic model that defines a posterior over the latent variables such as a variational auto-encoder \cite{2013arXiv1312.6114K} may allow more efficient sampling by performing inversion once and finding the region of matching geological models in latent space. 

\section{Conclusions}
Inversion of subsurface geological structures from acoustic reflection seismic data is a classical method to aid the understanding of the earth's interior. The inference of model parameters from measured acoustic properties is often performed in the very high-dimensional space of model properties leading to very CPU-intensive optimization \cite{akcelik2003high}.

We apply a method that combines a generative model of geological structures efficiently parameterized by a lower-dimensional set of latent variables, with a numerical solution of the acoustic inverse problem for seismic inversion using the adjoint method. Leveraging the adjoint of the studied partial differential equation we deduce gradients that are consequently used to sample from the posterior over the latent variables given the mismatch of the observed seismic data by following an approximate MALA scheme \cite{2016arXiv161200005N}.

While the proposed application was illustrated on a very simple geophysical inversion this method may find use in other domains where spatial property models control the evolution of physical systems such as in the flow of fluids in porous media or materials science. The combination of a deep generative model parameterized by a lower-dimensional set of latent variables and gradients obtained by the adjoint method may lead to new efficient techniques for solving high-dimensional inverse problems.
\section*{Acknowledgments}
O. Dubrule would like to thank Total S.A. for seconding him as visiting professor at Imperial College London.
\bibliography{example_paper}
\bibliographystyle{icml2018}
\newpage
\appendix
\section{Appendix}
\subsection{Generative Model Network Architectures}\label{sec:appendix}
\begin{table}[ht]
         \caption{\label{tab:gan_architecture}Generator and discriminator network architectures used to create synthetic geological structures. Binary indicators of geological facies and corresponding p-wave velocities are represented by a bi-variate Gaussian distribution and a hyperbolic tangent activation function was used to represent the two families of properties. Rock density shows a Gaussian distribution. A soft-plus activation function ($f(x)=\frac{1}{\beta}\log(1+exp(\beta~x)),~\beta=1$) was used to ensure positive numeric values of density. Notation for convolutional layers: LayerType(Number of filters),  k=kernel size, s=stride, p=padding. BN=BatchNorm, PS=PixelShuffle}
          \begin{subtable}{.45\textwidth}
              \centering
              {\begin{tabular}[t]{c}
                  \\
                  \toprule
	Latent Variables $z \in \mathbf{R}^{50 \times 1 \times 2}$ \\
                  \midrule
                  Conv2D(512)k3s1p1, BN, ReLU, PSx2 \\
                  \midrule
                  Conv2D(256)k3s1p1, BN, ReLU, PSx2 \\
                  \midrule
                  Conv2D(128)k3s1p1, BN, ReLU, PSx2 \\
                  \midrule
                  Conv2D(64)k3s1p1, BN, ReLU, PSx2 \\
                  \midrule
                  Conv2D(64)k3s1p1, BN, ReLU, PSx2 \\
                  \midrule
                  Conv2D(64)k3s1p1, BN, ReLU, PSx2 \\
                  \midrule
                  Conv2D(3)k3s1p1\\
                  \midrule 
                  Tanh (0,1) | Softplus (2) \\
                  \bottomrule
              \end{tabular}}
              \caption{\label{tab:generator} Multi-Channel Generator}
          \end{subtable}
          \hfill
          \begin{subtable}{.45\textwidth}
              \centering
              {\begin{tabular}[t]{c}
              \\
                  \toprule
	Geological Properties $m \in \mathbf{R}^{3 \times 64 \times 128}$ \\
                  \midrule
                  Conv2D(64)k5s2p2, ReLU \\
                  \midrule
                  Conv2D(64)k5s2p1, ReLU \\
                  \midrule
                  Conv2D(128)k3s2p1, ReLU \\
                  \midrule
                  Conv2D(256)k3s2p1, ReLU \\
                  \midrule
                  Conv2D(512)k3s2p1, ReLU \\
                  \midrule
                  Conv2D(512)k3s2p1, ReLU \\
                  \midrule
                  Conv2D(1)k3s1p1, ReLU \\
                  \bottomrule
              \end{tabular}}
              \caption{\label{tab:discriminator} Discriminator for multi-channel GAN}
          \end{subtable}
          \hspace{2mm}
\end{table}
\clearpage
\onecolumn
\subsection{Samples obtained by optimization in latent space}
\begin{figure*}[!h]
\vskip 0.2in
\begin{center}
\centerline{\includegraphics[width=\textwidth]{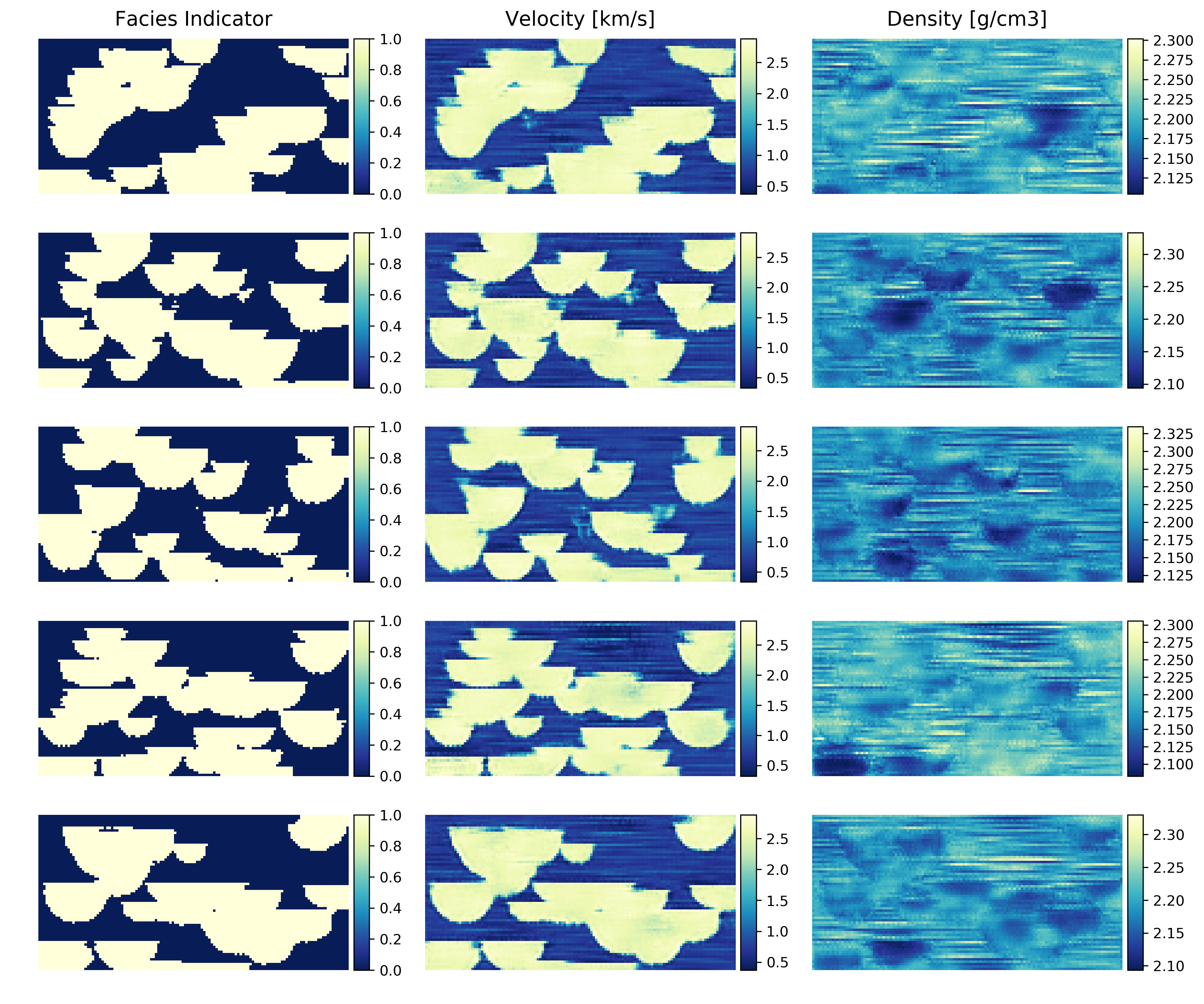}}
\caption{Samples obtained from latent space optimization with 2 acoustic sources.}
\label{fig:models_2_10}
\end{center}
\vskip -0.2in
\end{figure*}

\begin{figure*}[ht]
\vskip 0.2in
\begin{center}
\centerline{\includegraphics[width=\textwidth]{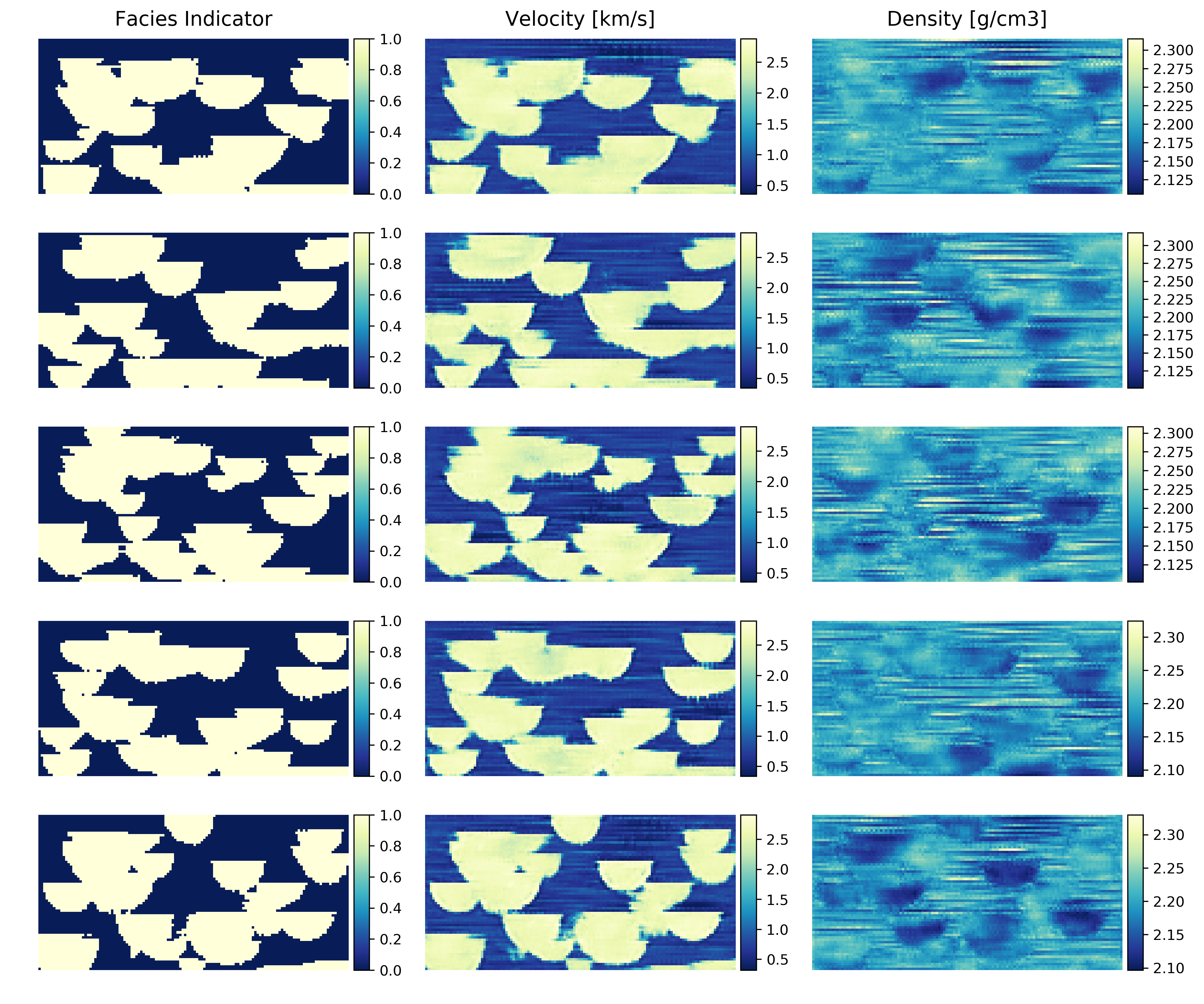}}
\caption{Samples obtained from latent space optimization with 3 acoustic sources.}
\label{fig:models_3_10}
\end{center}
\vskip -0.2in
\end{figure*} 
\FloatBarrier

\begin{figure*}[ht]
\vskip 0.2in
\begin{center}
\centerline{\includegraphics[width=\textwidth]{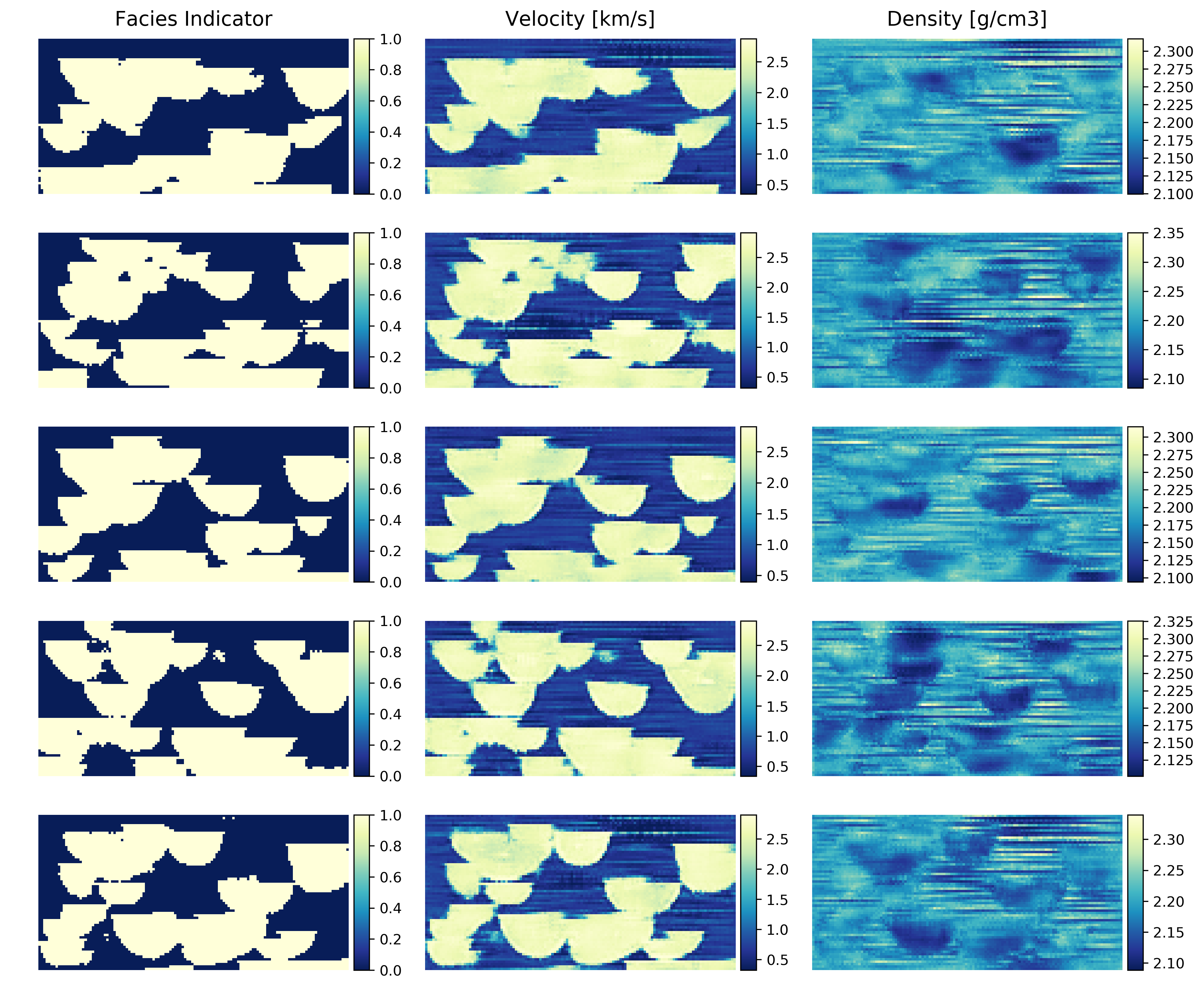}}
\caption{Samples obtained from latent space optimization with 9 acoustic sources.}
\label{fig:models_9_10}
\end{center}
\vskip -0.2in
\end{figure*} 
\FloatBarrier

\begin{figure*}[ht]
\vskip 0.2in
\begin{center}
\centerline{\includegraphics[width=\textwidth]{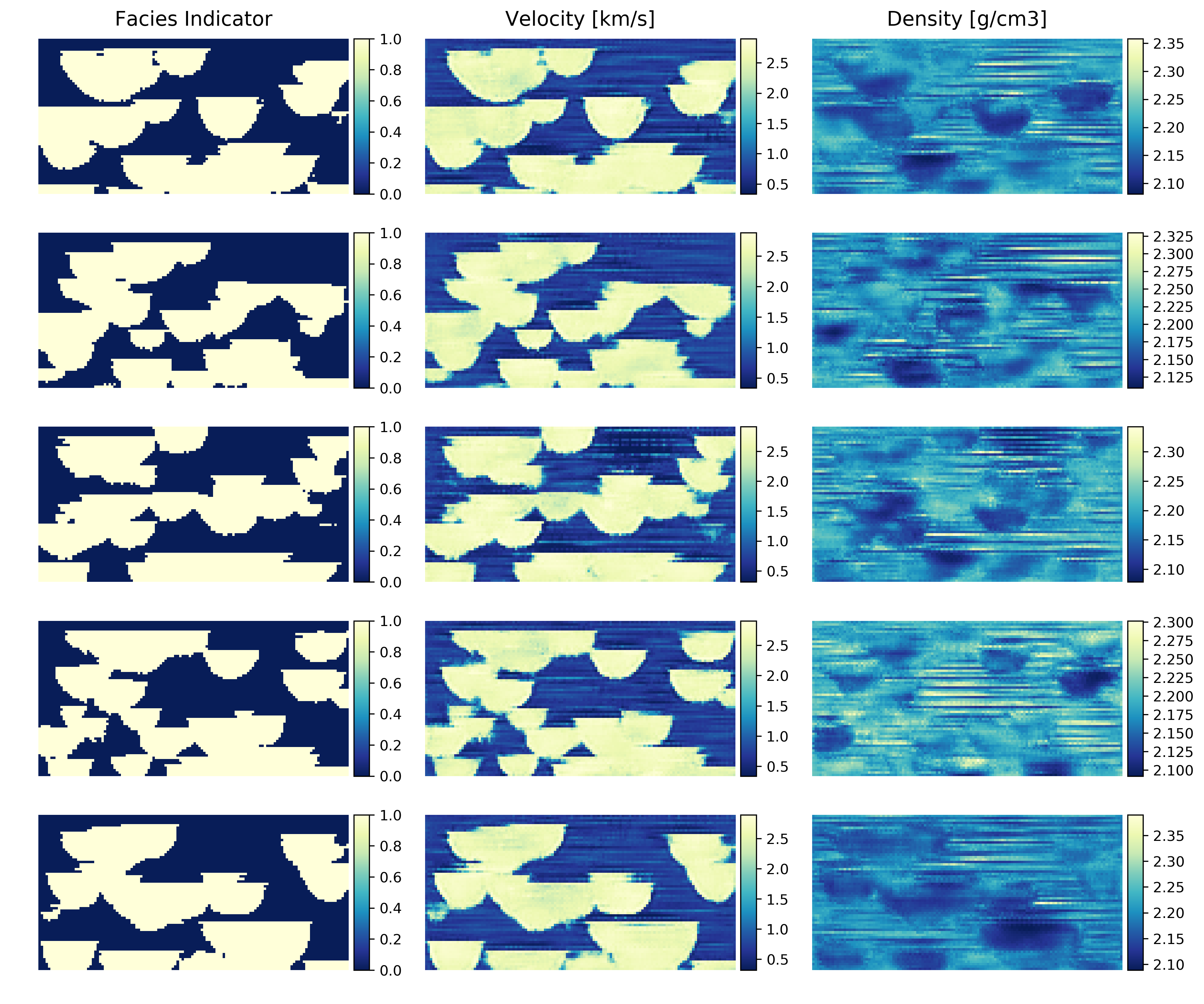}}
\caption{Samples obtained from latent space optimization with 27 acoustic sources.}
\label{fig:models_27_10}
\end{center}
\vskip -0.2in
\end{figure*} 
\FloatBarrier

\begin{figure*}[ht]
\vskip 0.2in
\begin{center}
\centerline{\includegraphics[width=\textwidth]{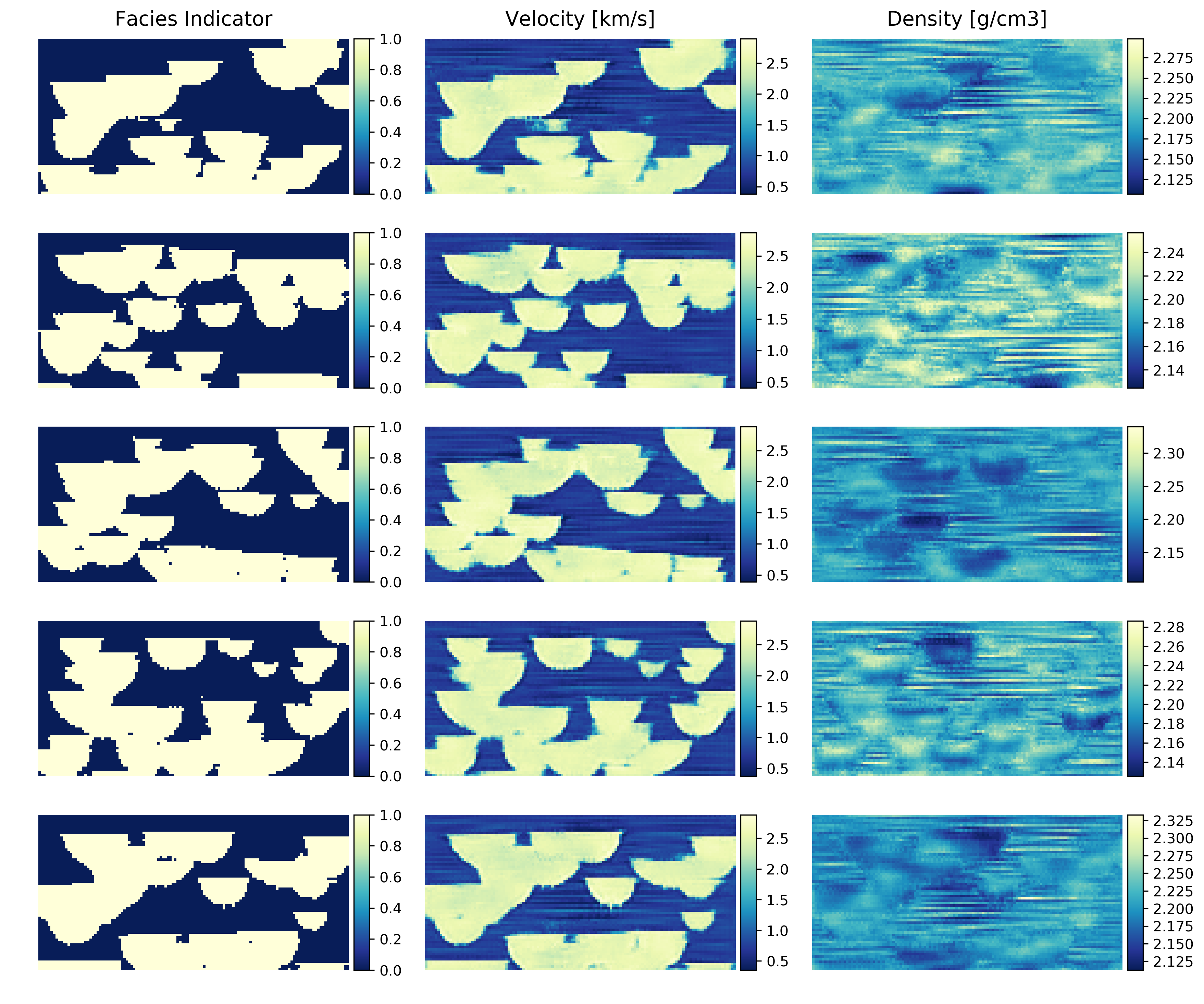}}
\caption{Samples obtained from latent space optimization with 2 acoustic sources and one bore-hole.}
\label{fig:models_2_10w}
\end{center}
\vskip -0.2in
\end{figure*} 

\end{document}